\documentstyle[prl,aps,epsf,multicol]{revtex}
\begin{document}
\draft
\tighten

\title{Two new topologically ordered glass phases of smectics confined in
 anisotropic random media}

\author{Brad Jacobsen$^1$, Karl Saunders$^2$, Leo Radzihovsky$^1$, 
John Toner$^2$}
\address{$^1$ Department of Physics, University of Colorado,
Boulder, CO 80309}
\address{$^2$ Dept. of Physics,
Materials Science Inst., and Inst. of Theoretical Science, University
of Oregon, Eugene, OR 97403}

\date{\today}
\maketitle
\begin{abstract}

Smectics in {\em strained} aerogel exhibit two new glassy phases. The
strain both ensures the stability of these phases and determines their
nature. One type of strain induces an ``XY Bragg glass'', while the
other creates a novel, triaxially anisotropic ``m=1 Bragg glass''. The
latter exhibits anomalous elasticity, characterized by exponents that
we calculate to high precision. We predict the phase diagram for the
system, and numerous other experimentally observable scaling laws.

\end{abstract}

\begin{multicols}{2}
\narrowtext

Liquid crystals confined in random porous structures, have become a
subject of considerable interest.\cite{review} A recent theoretical
study unambiguously demonstrated that conventional (quasi-)
long-ranged smectic order is impossible in 3d in the presence of (even
arbitrarily weak) quenched pinning imposed by these random structures,
e.g., aerogel.\cite{RTpr} It was proposed that a positionally
disordered but topologically ordered ``smectic Bragg glass'' (SBG)
phase would become the new thermodynamically distinct low-temperature
phase in these smectic systems. However, for quenched random {\em
isotropic} structures it was impossible to make a compelling
theoretical argument for the stability of such a glass phase.

\begin{figure}[bth]
\centering \setlength{\unitlength}{1mm}
\begin{picture}(30,27)(0,0)
\put(-46,-89){\begin{picture}(25,40)(0,0) 
\includegraphics{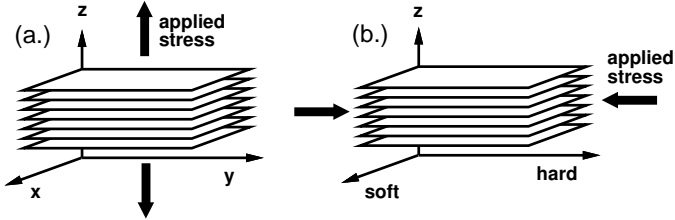}
\end{picture}}
\end{picture}
\caption{(a) Stretch along the ${\bf {\hat z}}$ direction.  (b)
Compression along $\perp$ direction.}
\label{squeeze_stretch}
\end{figure}

\begin{figure}[bth]
\centering \setlength{\unitlength}{1mm}
\begin{picture}(25,20)(0,0)
\put(-55,-90){\begin{picture}(25,45)(0,0) 
\includegraphics{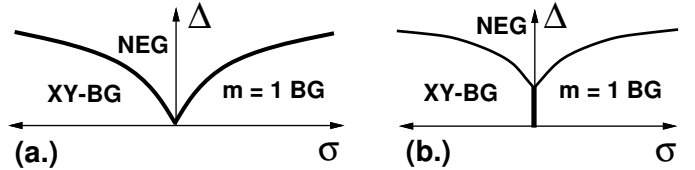}
\end{picture}}
\end{picture}
\caption{Two possible phase diagram topologies, depending on whether
S-BG is stable for isotropic confinement ($\sigma=0$).}
\label{phase_diagram}
\end{figure}

In this Letter, we show that we {\it can} make such a compelling
argument for smectics in a {\em uniaxially strained} aerogel, which
{\it certainly} exhibit two types of low-T BG phases, that are
thermodynamically distinct from the high-T nematic (or perhaps
``nematic elastic glass'' (NEG)) and isotropic liquid phases. For
parallel nematogen-surface alignment (assumed throughout), a stretch
(Fig.\ref{squeeze_stretch}a) of the aerogel will lead to an ``XY-BG''
in the isotropic universality class of randomly pinned vortex
lattices, CDW's, and random field XY magnets (RFXY),\cite{BraggGlass}
while a compression (Fig.\ref{squeeze_stretch}b) will lead to a novel
``$m = 1$ BG'', with triaxially
anisotropic scaling, that should be similar to that of a discotic in
{\em isotropic} aerogel\cite{SJRT}. For homeotropic alignment, the
phases reverse with respect to stretch and compression.
\vspace{-3mm}
We predict two possible low, constant-T phase diagrams, depending on
whether the SBG is stable (Fig.\ref{phase_diagram}b) or not
(Fig.\ref{phase_diagram}a).  Recent experiments\cite{BCRTscience}
suggest the former possibility.  The locii of the phase boundaries in
Fig.\ref{phase_diagram}a, for small strain, $\sigma$, are {\em
universal} and satisfy
\begin{equation}
\Delta(\sigma)\propto(K^3B)^{1/2}(\sigma/B)^\rho , \label{pb}
\end{equation}
where $\sigma$ is proportional to the uniaxial stress applied to the
aerogel fibers, $\Delta$ is a measure of the tilt disorder, $B$ and
$K$ are bulk smectic elastic moduli, and $\rho$ a universal exponent
expressible in terms of anomalous elasticity exponents
$\tilde{\eta}_B$ and $\tilde{\eta}_K$ for {\em unstrained}
aerogel. Our best estimate is $\rho\approx1/3$ in 3d.\cite{RTpr}

Our model of the smectic in aerogel treats the local smectic layer
displacement $u({\bf r})$ and the local nematic director $\hat{\bf
n}({\bf r})$ as the only important fluctuating quantities, ignoring
fluctuations in the magnitude $|\psi|$ of the smectic order parameter
$\psi=|\psi| e^{i q_o u({\bf r})}$ about its mean $|\psi_o|$.
%
The important effects of the aerogel are {\em completely} described by
only two disorder types. One is the random-field translational
disorder $\delta H_{rf}={\cal R}e \int \!  d^d{\bf r}|\psi_o| V({\bf
r})e^{i q_o u({\bf r})}$, coupling to $u({\bf r})$, with $V({\bf r})$
a complex random potential which at long scales can be accurately
represented as zero-mean and {\em short-ranged}\cite{RTpr} Gaussian
statistics with $\overline{V({\bf r})V^*({\bf r'})} =
\tilde{\Delta}_V
\delta^d({\bf r}-{\bf r'})$. The other type of disorder is the
random-tilt orientational disorder given by $\delta H_{t}=-\int \!
d^d{\bf r} ({\bf g}({\bf r})\cdot{\hat{\bf n}})^2$, describing the
tendency of nematogens $\hat{\bf n}({\bf r})$ to align along the
local aerogel strand directed along ${\bf g}({\bf r})$, and at
long-scales is completely described by {\em short-ranged}\cite{RTpr}
correlations $\overline{g_i({\bf r})g_j({\bf r'})} =
1/2(\sqrt{\Delta}\delta_{ij}-\gamma {e_i} {e_j})\delta^d({\bf r}-{\bf
r'})$, with $\hat{\bf e}$ the uniaxial direction (i.e., the axis of the strain
applied to the aerogel). In the above $\tilde{\Delta}_V = \Gamma_u
(a_f/L_f)^{d-d_F}(1/(a_f q_o))$, $\Delta = \Gamma_n
(a_f/L_f)^{d-d_F}$, $d_f$ is the aerogel's fractal dimension for scales
$a_f<r<L_f$, and $\Gamma_u$, $\Gamma_n$, $\gamma$ are phenomenological
parameters, with $\gamma$ the anisotropy parameter which at small
strains is proportional to $\sqrt{\Delta}$ and the stress $\sigma$
applied to the aerogel. $\gamma < 0$ for a stretch illustrated in
Fig.\ref{squeeze_stretch}a.


Assuming (as we'll verify a posteriori) that fluctuations in $\hat{\bf
n}$ from a perfect alignment with the smectic layer normal (taken
along $\hat{\bf z}$) are small, allows us to integrate $\hat{\bf n}$
out of the partition function, with the only effect of replacing
$\delta{\bf n}\rightarrow \bbox{\nabla}_\perp u$.\cite{RTpr} The
resulting Hamiltonian is given by
\begin{eqnarray}
H&=&\int_{\bf r} \bigg[ {K\over2}(\bbox{\nabla}_\perp^2 u)^2 +
{B\over 2}\Big(\partial_z u-{1\over
2}(\bbox{\nabla}_\perp u)^2\Big)^2\nonumber\\ 
&+&(g_z({\bf r})\bbox{\nabla}_\perp u)^2
- ({\bf g}({\bf r}) \cdot \bbox{\nabla}_\perp u)^2 -
2 g_z({\bf r}){\bf g}({\bf r}) \cdot \bbox{\nabla}_\perp u\nonumber\\
&-&|\psi_o|{\cal R}e\left\{V({\bf r}) e^{i q_o u({\bf r})}\right\}\bigg]\;. 
\label{H_XY}
\end{eqnarray}
The form of the anharmonic elastic terms is dictated by the underlying
invariance of the {\em bulk} smectic phase under rotations about any
axis lying in the ${\bf r}_\perp$-plane.

After introducing $n$ replica fields $u_\alpha$ and integrating out
the disorder, we obtain the Hamiltonian whose form strongly depends on
the type of uniaxial strain. {\em Stretching} the aerogel strands
will cause the layer normal, $\hat{\bf z}$, to align with $\hat{\bf e}$
(Fig.\ref{squeeze_stretch}a). Smectics
confined inside this structure, to harmonic order in elasticity (with
elastic anharmonicity irrelevant) are described by
\begin{eqnarray}
H_{XY} &=& {1\over2} \int_{\bf r} \sum_{\alpha=1}^n \bigg[K
(\bbox{\nabla}_\perp^2 u_\alpha)^2 \hspace{-1mm}+ B (\partial_z
u_\alpha)^2
\hspace{-1mm}+
|{\gamma}|(\bbox{\nabla}_\perp u_\alpha)^2 \bigg] \nonumber\\ &-&
{1\over2T} \int_r \sum_{\alpha,\beta=1}^n \bigg[ (\Delta +
\sqrt{\Delta} |\gamma|)(\bbox{\nabla_\perp}u_\alpha) \cdot
(\bbox{\nabla_\perp}u_\beta) \nonumber\\ 
&+& \Delta_V\cos[q_o(u_\alpha-u_\beta)] \bigg]\;, 
\label{H_XYr}
\end{eqnarray}
where $\Delta_V=2|\psi_o|^2\tilde{\Delta}_V$. At scales smaller than a
crossover scale $\xi_\perp^c$ (to be defined below), the behavior is
that of a smectic pinned by {\em isotropic} unstrained
aerogel.\cite{RTpr} On longer scales, however, such anisotropically
pinned smectic crosses over to {\em isotropic} scaling behavior of the
RFXY model. We therefore predict that smectics pinned by such
anisotropic weak disorder will exhibit the XY-BG phase, with its
universal {\em disorder}-induced logarithmic layer wandering
character, $\overline{\langle u^2({\bf
r})\rangle}=C(d)(\ln{r})/q_o^2$.\cite{BraggGlass} However, unlike $3d$
{\em bulk} smectics, which show the famous Landau-Peierls {\em
thermally}-driven $\ln{r}$ fluctuations, here $C(d)$ is {\it
universal}, the logarithm persists in all $2<d<4$, and smectic layers
are pinned. The immediate consequence is that X-ray scattering will
exhibit real-space power-law decay $\overline{\langle\rho_G({\bf
r})\rho_{-G}({\bf 0})\rangle} \propto r^{-\eta(G)}$ with a {\em
universal} $\eta(G)$ exponent ($G=m q_o$).

If, instead, the aerogel is uniaxially {\em compressed}, i.e.,
$\gamma>0$, we expect that one of the (previously soft) ${\bf
r}_\perp$ smectic axes ($x$ or $y$) will orient along the axis of
compression $\hat{\bf e}_h$ (Fig.\ref{squeeze_stretch}b). We denote
this $\hat{\bf e}_h$-directed axis as `'{\em hard}'' ({\em h}), and call
the other $\perp$ axis, orthogonal to $\hat{\bf e}_h$, '`{\em soft}''
({\em s}) axis, i.e., ${\bf r}_\perp=({\bf r}_h,{\bf r}_s)$.  The
resulting effective Hamiltonian describing this system at long scales
is
\begin{eqnarray}
H_{m=1} &=& {1\over2} \int_{\bf r}\bigg[ \sum_{\alpha=1}^n K
(\bbox{\nabla}_\perp^2 u_\alpha)^2 + B\Big(\partial_z u_\alpha - {1\over
2}(\bbox{\nabla}_\perp u_\alpha)^2\Big)^2 \nonumber\\ 
&+& \gamma(\bbox{\nabla}_h u_\alpha)^2 - 
\sum_{\alpha,\beta=1}^n {\Delta \over T} 
(\bbox{\nabla}_\perp u_\alpha)\cdot(\bbox{\nabla}_\perp u_\beta)\bigg]\;,
\label{H_m=1r}
\end{eqnarray}
where we have neglected the positional random-field disorder,
$\Delta_V$, which can be shown to be subdominant at long length
scales.\cite{SJRT} $H_{m=1}$, Eq.\ref{H_m=1r}, implies that the
noninteracting propagator $G_{\alpha\beta}({\bf q})\equiv
V^{-1}\langle u_\alpha({\bf q})u_\beta(-{\bf q})\rangle_0= T G({\bf
q})\delta_{\alpha\beta}+ \Delta q_\perp^2 G({\bf q})^2$, with $G({\bf
q})\equiv 1/(K q_\perp^4+\gamma q_h^2+B q_z^2)$. As usual, at long
length scales, the disorder ($\Delta$) contribution to layer roughness
dominates over the thermal ($T$) part of $G_{\alpha\beta}({\bf q})$.

We first note that for vanishing strain $\gamma\gtrsim 0$, or
equivalently at very short length scales, $H_{m=1}$ and the
corresponding propagator reduce to those characterizing a smectic in
{\em unstrained isotropic} aerogel.\cite{RTpr} The asymptotic long
scale behavior of the full model described by $H_{m=1}$ is reached via
two independent crossovers from the Gaussian, unstrained fixed point,
during which the aerogel anisotropy $\gamma$, and the nonlinear
elasticity, respectively become important. The qualitative form of
this crossover is determined by the relative magnitudes of the
corresponding bare couplings. For sufficiently weak strain
$(\gamma<\gamma_c$), the elastic anharmonicity becomes important first
and this occurs at a crossover length scale
$\tilde{\xi}_\perp^{NL}\propto({K^{5/2}\over {B^{1/2}
\Delta}})^{1\over5-d}$ determined by the smectic in {\em unstrained}
aerogel.\cite{RTpr} In this case the system first crosses over from
the unstrained Gaussian to {\em unstrained} anomalous fixed point. The
final crossover to asymptotic {\em strained} anomalous behavior takes
place within the anomalously elastic smectic described by the
wavevector-dependent elastic constants\cite{RTpr} and occurs at ${\bf
q}_\perp$ such that $\tilde{K}({\bf q}_\perp)q_\perp^4\approx\gamma
q_\perp^2$, with $\tilde{K}({\bf q}_\perp)$ calculated in
Ref.\onlinecite{RTpr}, i.e. at $\xi_\perp^c\approx
[K/(\gamma(\tilde{\xi}_\perp^{NL}))^{\tilde{\eta}_K}]^{1/(2-\tilde{\eta}_K)}$
(we use tilde symbol for exponents for {\em isotropic} disorder).

For the remainder of this paper we will focus on the other
crossover scenario in which the strain $\gamma$ is sufficiently large
$(\gamma>\gamma_c$) that the crossover from Gaussian unstrained
to Gaussian strained elasticity takes place at
$\xi_\perp^c=\sqrt{K/\gamma}$, {\em before} elastic nonlinearities
become important. The critical value of $\gamma$ that delineates
between these two crossover scenarios is
$\gamma_c=K/(\tilde{\xi}_\perp^{NL})^2$.

For $\gamma>\gamma_c$, on scales longer than
$\xi_\perp^c=\sqrt{K/\gamma}$ the effective Hamiltonian (and the
propagator $G$ derived from it) is identical to that given in
Eq.\ref{H_m=1r}, but with all ${\bbox\nabla}_\perp$ replaced by
${\bbox\nabla}_s$, with ${\bf r}_s$ a subset of ${\bf r}_\perp$ axes
remaining soft even in the presence of aerogel anisotropy.  Our goal
then is to assess the role of elastic nonlinearities, at this new
{\em strained} Gaussian fixed point, which become important beyond 
an even longer nonlinear crossover length scale $\xi_s^{NL}$ (along
the `soft' direction).

The scale $\xi_s^{NL}$ can be determined from a simple perturbation
theory in these nonlinear couplings of $H_{m=1}$, and is the length at
which the effects of anharmonic elastic terms become significant.  For
example, the diagrammatic correction to the bulk modulus $B$, due to
these elastic nonlinearities, is given by
\begin{mathletters}
\begin{eqnarray}
\delta B(L)&=&-{B^2\over2}\int_{\bf q}^>\left[T G({\bf q})^2
+2\Delta q_s^2 G({\bf q})^3\right]q_s^4\;,\\
\label{deltaBa}
&\approx&{-C_{d-1}\beta_{d-1}\over 2\pi(7-2d)} \Delta \bigg({B^3
\over\gamma^{d-2}K^{7-d}}\bigg)^{1/2}L^{7-2d}\;,
\label{deltaBb}
\end{eqnarray}
\end{mathletters}
where in the above we have kept only the dominant disorder infrared divergent
contribution, cutoff these long scale divergences by $q_s>1/L$, and
analytically continued to arbitrary dimension $d$, with a single
smectic ordering coordinate $z$, a single soft coordinate $r_s$, and
$d-2$ hard axes with coordinate ${\bf r}_h$.  The constant
$C_d=2\pi^{d/2}/\left((2\pi)^d\Gamma(d/2)\right)$ and
$\beta_d=\Gamma(d/2)\Gamma(3-d/2)/2$. For $d<d_{uc} = 7/2$, the
corrections to $B$, Eq.\ref{deltaBb}, grow with cutoff $L$ and become
significant for scales $L>\xi_s^{NL}$, such that $|\delta
B(\xi_s^{NL})|=B$, signaling the breakdown of conventional harmonic
elasticity. We find
\begin{equation}
\xi_s^{NL}=\bigg({2\pi(7-2d)K^{(7-d)/2}\;\gamma^{(d-2)/2}
\over C_{d-1}\beta_{d-1}\;\Delta\; B^{1/2}}\bigg)^{1/(7-2d)} 
\label{xi_sNL}
\end{equation}
The corresponding lengths along the $z$ and $h$ axes are given by
$\xi_z^{NL}=(\xi_s^{NL})^2/\lambda_B$ and
$\xi_h^{NL}=(\xi_s^{NL})^2/\lambda_\gamma$, where
$\lambda_B\equiv(K/B)^{1/2}$ and $\lambda_\gamma
\equiv(K/\gamma)^{1/2}$. Identical crossover lengths scales are
obtained if one instead studies perturbative corrections to $K$ or
$\Delta$.

To go beyond these crossover length scales $\xi_{z,h,s}^{NL}$ we use
the renormalization group (RG), which consists of integrating out
short-scale modes, perturbatively in elastic nonlinearities, and
rescaling the lengths and long wavelength part of the fields with
$r_s=r_s'e^{\ell}$, $r_h=r_h'e^{\omega_h\ell}$, $z=z'e^{\omega_z\ell}$
and $u_\alpha({\bf r})= e^{\chi\ell}u_\alpha'({\bf r'})$, so as to
restore the uv cutoff back to $\Lambda\sim 1/a$.  The underlying
rotational invariance ensures that the graphical corrections preserve
the rotationally invariant operator $\big(\partial_z u_\alpha -
{1\over2}(\bbox{\nabla}_s u_\alpha)^2\big)$, renormalizing it as a
whole. It is therefore convenient (but not necessary) to choose the
dimensional rescaling that also preserves this operator; the
appropriate choice is $\chi=2-\omega_z$.

Using the above defined analytical continuation in $d$, RG
to one-loop order, gives the following flow
equations
\begin{mathletters}
\begin{eqnarray}
\frac{d B(\ell)}{d\ell}&=&
\Big(5+(d-2)\omega_h-3\omega_z-{3g\over{32\sqrt2}}\Big)B
\label{Bflow}\;, \\
\frac{d K(\ell)}{d\ell}&=&
\Big(1+(d-2)\omega_h-\omega_z+{g\over{8\sqrt2}}\Big)K
\label{Kflow}\;, \\
\frac{d(\Delta/T)(\ell)}{d\ell} 
&=& \Big(3+(d-2)\omega_h-\omega_z+{g\over{32\sqrt2}}\Big)(\Delta/T)
\label{Deltaflow}\;,
\end{eqnarray}
\end{mathletters}
where we've defined a dimensionless measure of disorder $g
\equiv \Delta(B/(K^{7-d}\gamma^{d-2})^{1/2}C_{d-1}\Lambda^{2d-7}$, which flows according to
\begin{eqnarray}
\frac{d g(\ell)}{d\ell}&=&2\epsilon g -
{15\over{64\sqrt2}} {g}^2\,\label{g2flow}
\end{eqnarray}
with $\epsilon\equiv7/2-d$. Because all relevant anharmonic terms in
$H_{m=1}$ appear with ${\bbox\nabla}_s$, there are no graphs
correcting $\gamma$ and therefore no anomalous $\gamma$ elasticity
to all orders.  As required, the flow of $g$ is independent of the
arbitrary choice of the anisotropy rescaling exponents $\omega_h$ and
$\omega_z$. The growth of $g$ for $d<d_{uc}=7/2$ is an
indication that the long scale properties of our system, even at a finite
temperature $T$, are dominated by disorder. The eventual termination of
this flow at a nontrivial, glassy $T=0$ fixed point ${g}^*=\epsilon
128\sqrt2/15$, leads to strong disorder-generated power-law anomalous
elasticity.

One consequence of the anomalous elasticity is that the long-scale
elastic constants $K$, $B$, and disorder variance $\Delta$ become
wavevector dependent.
%
\begin{mathletters}
\begin{eqnarray}
K({\bf k})&=&K k_s^{-\eta_K}
f_K\left({k_h/k_s^{\zeta_h}}, {k_z/k_s^{\zeta_z}}\right)\;,
\label{K}\\ 
B({\bf k})&=&B k_s^{\eta_B}
f_B\left({k_h/k_s^{\zeta_h}}, {k_z/k_s^{\zeta_z}}\right)\;,
\label{B}\\ 
\Delta({\bf k})&=&\Delta k_s^{-\eta_{\Delta}}
f_{\Delta}\left({k_h/k_s^{\zeta_h}}, {k_z/k_s^{\zeta_z}}\right)\;,
\label{Delta}
\end{eqnarray}
\label{anom_elasticity}
\end{mathletters}
$\gamma({\bf k})=\gamma$, with the anisotropy exponents $\zeta_z\equiv
2-(\eta_B+\eta_K)/2$ and $\zeta_h\equiv 2-\eta_K/2$. The exponents obey 
\begin{equation}
7-2d + \eta_\Delta={\eta_B \over 2} + {7-d\over
2}\eta_K\;,
\label{WIh}
\end{equation}
exactly, due to the underlying {\em exact} rotational invariance of
Eq.\ref{H_m=1r} about $\hat{\bf e}_h$. To leading order in $\epsilon=7/2-d$,
%
%
$\eta_K=g^*/8\sqrt2=16\epsilon/15=8/15$,
$\eta_B= 3g^*/{32\sqrt2}=12\epsilon/15=2/5$, and
$\eta_{\Delta}= g^*/{32\sqrt2}=2\epsilon/15=2/15$,
the last equalities holding in $d=3$ ($\epsilon=1/2$).
Since $\epsilon=1/2$ is quite small,
we expect these exponents to be quantitatively accurate.  

The RG $\epsilon=7/2-d$ expansion treatment presented above is nicely
complemented by an $\hat{\epsilon}\equiv 4-d$ expansion arising from a
different analytical continuation to $d$ dimensions, in which there
are $d-2$ soft coordinates ${\bf r}_s$ and only a single hard axis.
The corresponding exponents are given by
$\hat{\eta}_K=3\hat{\epsilon}/8=3/8$,
$\hat{\eta}_B=3\hat{\epsilon}/4=3/4$, and $\hat{\eta}_{\Delta}
=\hat{\epsilon}/8=1/8$, with good agreement in $d=3$ (except for
$\eta_B$) with the $\epsilon=7/2-d$ expansion results. The exact
exponent relation for the $\hat{\epsilon}=4-d$ expansion is given by
$4-d+\hat{\eta}_\Delta=\hat{\eta}_B/2 + 2\hat{\eta}_K$ and
reassuringly agrees with Eq.\ref{WIh} in $d=3$.

Further accuracy can be gained by weighted-averaging of
the $7/2-d$ and $4-d$ expansions, according to: $\eta_{K,\Delta}
\rightarrow(4\eta_{K,\Delta}+\hat{\eta}_{K,\Delta}^{s})/5$.  The
factor of $4$ reflects the higher accuracy of the $7/2-d=\epsilon$
expansion.  The prediction for
$\eta_B$ is then made using the exact scaling relation, giving, in
$d=3$,
$\eta_K=0.50$, $\eta_B=0.26$, and $\eta_{\Delta}=0.13$.
%
%

We now study the translational and orientational order in the presence
of this strong disorder-driven anomalous elasticity.  The former is
characterized by the growth of smectic layer roughness with e.g., $r_s$
\begin{eqnarray}
C(r_s)&\equiv&\overline{\langle\left(u(r_s,0_h,0_z)-
u(0_s,0_h,0_z)\right)^2\rangle}\;,\nonumber\\
&=&\int{d^{3}k\over(2\pi)^{3}} 2(1-\cos(k_s r_s))\Delta({\bf k})
k_\perp^2G^2({\bf k})
\;,\label{Cs}
\end{eqnarray}
from which the translational correlation length $\xi_s^{X}$ (the
inverse of the X-ray diffraction peak width), can be computed via the
condition $C(r_s=\xi_s^{X})\equiv a^2$, with $a$ the smectic layer
spacing. $\xi_s^{X}$ is determined by the relative order of
many crossover length scales. For $\xi_s^X < \xi^c_\perp$, $\xi_s^{X}$
is identical to that due to {\em isotropic} disorder.\cite{RTpr} In
the opposite regime, $\xi_s^X>\xi^c_\perp$, there are three
possibilities, depending on whether the anomalous elasticity sets in
before or after the layer roughenness reaches $a$, and whether the
isotropic-to-anisotropic crossover takes place near the harmonic or
the anomalous elastic fixed point:
\begin{equation}
\xi_s^{X}=\cases{\xi_s^{NL}({a\over\lambda_B})^2, &
$\xi^c_\perp<\xi_s^{X}<\xi^{NL}_s$, $\gamma>\gamma_c$ \cr
\xi_s^{NL}({a\over\lambda_B})^{{2\over(\eta_B+\eta_K)}}, &
$\xi^c_\perp<\xi^{NL}_s<\xi_s^{X}$, $\gamma>\gamma_c$ \cr 
\xi^c_\perp({a\over\lambda_B})^{{2\over(\eta_B+\eta_K)}}\times \cr 
(\xi^c_\perp/\tilde{\xi}_\perp^{NL})^{{\tilde{\eta}_B+\tilde{\eta}_K
\over\eta_B+\eta_K}}, &
$\tilde{\xi}^{NL}_\perp<\xi^c_\perp<\xi_s^{X}$, $\gamma<\gamma_c$}
\label{xiX}
\end{equation}
From these correlation lengths we see that it is the ratio
$a/\lambda_B$ which determines whether $\xi_s^{X}$ lies in a length
scale regime in which anharmonic effects are important. For small
$B$, $\lambda_B\gg a$ and anharmonic effects are unimportant. Note
also that in the strained length scale regime, $\xi_s^{X}$ will depend
on $B,K,\Delta$ and $\gamma$. Thus, one could test the predictions of
Eq.\ref{xiX} by measuring the dependence of $\xi_s^{X}$ on the
strength of compression (i.e., $\gamma$) which could be adjusted
directly. In all length scale regimes, the X-ray correlation length is
finite even as $T \rightarrow 0$, signaling the destruction of the
conventional (quasi-) long-ranged translational smectic order.

As emphasized in Ref.\onlinecite{RTpr}, this lack of translational
order does {\em not} imply that the low temperature phase replacing
the smectic is simply a nematic (or ``worse'', isotropic). Our
detailed calculations\cite{SJRT} indicate that, in fact, despite the
lack of the (quasi-) long-ranged smectic order dislocation loops
remain bound for weak {\em anisotropic} disorder, and therefore the
low temperature phase replacing the smectic must be distinct from the
nematic, separated from it by a thermodynamically sharp 
dislocation unbinding phase
transition. We call this putative low-temperature phase the ``$m=1$ -Bragg Glass''.

The stability of this exotic glass phase is contingent upon our
implicit assumption of long-ranged orientational (nematic) order.
That this assumption is valid can be easily seen by computing
$\overline{\langle|\delta{\bf
n}|^2\rangle}=\overline{\langle|{\bbox\nabla}u|^2\rangle}$, and taking into
account the wavevector dependent elastic moduli $K({\bf k})$ and
$B({\bf k})$ and disorder variance $\Delta({\bf k})$, as given by
Eqs.\ref{anom_elasticity}.  There are unstrained and strained
contributions to $\overline{\langle|\delta{\bf n}|^2\rangle}$, arising from modes
with $q_\perp>1/\xi_\perp^c$ and $q_\perp<1/\xi_\perp^c$,
respectively.
%
Using the corresponding anomalous exponents $\eta_K$, $\eta_B$,
$\eta_\Delta$,\cite{RTpr} in the
computation of the strained and unstrained parts, for finite strain
$\gamma$ and weak disorder $\Delta$, we indeed find long-ranged
orientational order. In the weak strain limit ($\gamma<\gamma_c$), the unstrained part
dominates, in 3d growing in a {\em universal} way with decreasing
strain $\gamma$ and increasing disorder $\Delta$ as
$(\Delta^\mu/\gamma^{\mu-1})$, where $\mu=\tilde{\eta}_B/(2-\tilde{\eta}_K)$.
Using Ref.\onlinecite{RTpr} we estimate $\mu$ to be $3/2$.
Since $\overline{\langle|\delta{\bf
n}|^2\rangle}$ can therefore get arbitrarily large at small $\gamma$
and large $\Delta$, we expect our system to be in the orientationally
disordered liquid phase in this range of parameters. On the other hand
for large $\gamma$ and small $\Delta$ the system will exhibit
long-ranged orientational order and as illustrated in
Fig.\ref{phase_diagram}a will therefore be  in the $m=1$ BG
phase. In analogy with the Lindemann
criterion for melting, the phase boundary is roughly determined by the
condition $\overline{\langle|\delta{\bf n}|^2\rangle}\approx O(1)$.
This leads to the phase boundary quoted in Eq.\ref{pb}, and
illustrated in Fig.\ref{phase_diagram}a.

On the other hand, rather than rely on the untrustworthy (in 3d)
$5-\epsilon$ expansion, which predicts no SBG for isotropic
($\sigma=0$) disorder\cite{RTpr}, we can infer the topology of the
phase diagram based on the preliminary experimental
evidence\cite{BCRTscience}, which suggests the stability of SBG for weak
isotropic disorder. This suggests that the $m=1$ BG extends all the
way down to vanishing strain, $\sigma>0$, as illustrated in
Fig.\ref{phase_diagram}b.

Light scattering, which measures director correlation
$\overline{\langle\delta n_i({\bf q})\delta n_j({\bf -q})\rangle}$
%
%
provides an independent means to test the predictions of the
theory. 
Finally, since anomalous elasticity also implies a nonlinear
stress-strain relation at arbitrarily weak stress, our predictions for
it can be independently probed in an a.c. acoustic experiment,
searching for an unusually large second harmonic response.

L.R. and B.J. acknowledge support by the NSF DMR-9625111, the MRSEC
DMR-9809555, and the A.P. Sloan and David and Lucile Packard
Foundations. J.T. and K.S. were supported by the NSF DMR-9634596.

\end{multicols}
\end{document}